\documentclass[twocolumn,superscriptaddress,showpacs,preprintnumbers,amsmath,amssymb,floatfix]{revtex4}
\usepackage{graphicx}

\begin{document}

\title{Computer Assembly of Cluster-Forming Amphiphilic Dendrimers}

\author{Bianca~M.~Mladek}
\affiliation{Center for Computational Materials Science and Institut f\"ur
  Theoretische Physik, Technische Universit\"at Wien, Wiedner Hauptstra{\ss}e
  8-10, A-1040 Wien, Austria}

\author{Gerhard~Kahl}
\affiliation{Center for Computational Materials Science and Institut f\"ur
  Theoretische Physik, Technische Universit\"at Wien, Wiedner Hauptstra{\ss}e
  8-10, A-1040 Wien, Austria}

\author{Christos~N.~Likos}
\affiliation{Institut f{\"u}r Theoretische Physik II: Weiche Materie,
Heinrich-Heine-Universit{\"a}t D{\"u}sseldorf,
Universit{\"a}tsstra{\ss}e 1, D-40225 D{\"u}sseldorf,
Germany}

\date{\today}

\begin{abstract}
Recent theoretical studies have
predicted a new clustering mechanism for soft matter
particles that interact via a certain kind of purely repulsive,
bounded potentials.  At sufficiently high densities, clusters of
overlapping particles are formed in the fluid, which upon 
further compression crystallize into cubic lattices with 
density-independent lattice constants.
In this work we show that amphiphilic dendrimers are suitable
colloids for the experimental realization of this phenomenon. Thereby,
we pave the way for the synthesis of such macromolecules, which form
the basis for a novel class of materials with unusual properties. 
\end{abstract}


\maketitle

Self-assembly is the key-expression that circumscribes the incredibly
rich wealth of ordered phases encountered in soft matter
systems. Starting from the face-center cubic arrangement assumed by
colloidal spheres at high concentrations
\cite{pusey:nature:86,gasser:science:01,henk:review:02} and
low-symmetry crystals formed by soft spheres \cite{primoz:prl:00},
they extend over to a variety of alloys seen for charged colloidal
mixtures \cite{leunissen:05} and to the gyroid phases assembled in
block copolymer solutions \cite{hajduk:94,matsen:98,ullal:04}. Recent
theoretical and computational advances have predicted a novel form of
self-assembly in soft matter, i.e., the formation of stable clusters
\cite{bmm:prl:06,cnl:jcp:07} encountered for {\it purely repulsive},
bounded effective potentials $\Phi_{\rm eff}(r)$. These 
clusters crystallize into 
cubic lattices at sufficiently high densities and all
temperatures 
\cite{bmm:prl:06,cnl:jcp:07}. This phenomenon 
bears significant consequences both from the
fundamental point of view \cite{cnl:jcp:07} and from the aspect
of the properties of the ensuing materials, e.g., their diffusion and
relaxation dynamics \cite{angel:prl:07}. 
Though thoroughly understood at the level of effective potentials,
the phenomenon begs the
question: {\it What kinds of particles display the class
of effective interactions giving rise to this phase behavior?}
In this
contribution we demonstrate that 
relatively
simple macromolecules can be designed to achieve this aim,
taking advantage of the
great flexibility offered by soft matter systems to manufacture new
materials.

The underlying theoretical concepts can be clearly stated as 
follows. 
Repulsion-induced aggregation which leads to ordered cluster phases
requires that the Fourier transform (FT) $\tilde\Phi_{\rm
eff}(k)$ of $\Phi_{\rm eff}(r)$ has negative parts for some values of
the wave-vector $k$. 
If, however, $\tilde\Phi_{\rm eff}(k) > 0$ for
all $k$, reentrant melting occurs instead \cite{cnl:pre:01}. 
A sufficient condition for
the former is that $\Phi''_{\rm eff}(r = 0) \geq
0$ \cite{cnl:jcp:07}. Useful archetypes of bounded interactions
\cite{bmm:prl:06,cnl:jcp:07} are the generalized exponential models of
index $n$ (GEM-$n$), where $\Phi_n(r) = \varepsilon
\exp[-(r/\sigma)^n]$, with $\varepsilon$ and $\sigma$ being some
energy- and length-scales: here, for $n > 2$ clustering takes place,
whereas for $n \leq 2$ reentrant melting occurs
\cite{lang:jpcm:00}. 

Searching for realizations of clustering-type potentials, we
concentrate on dendrimers, a choice motivated by their outstanding
properties.
They are characterized by a high degree of monodispersity and a
well-defined, highly branched internal structure; efficient 
dendrimer
assembly has been boosted by recent progress in synthetic techniques
\cite{antoni:07}. Fundamentally, they serve as tunable soft colloids
that allow for control of their effective interactions via changes in
chemical composition, bond length and generation number
\cite{ingo:jcp:04,ac:04}. For athermal dendrimers, $\Phi_{\rm eff}(r)$
has a Gaussian shape \cite{ingo:jcp:04} ($n=2$ in the
GEM-$n$ family), hinting
thus to the reentrant melting scenario. Closely linked with these
findings is the dense-core structure of these dendrimers, arising from
back-folding of the terminal groups \cite{ac:04,zook:prl:03}.

\begin{figure}
\begin{center}
\includegraphics[width=\columnwidth, clip=true, draft=false]
 {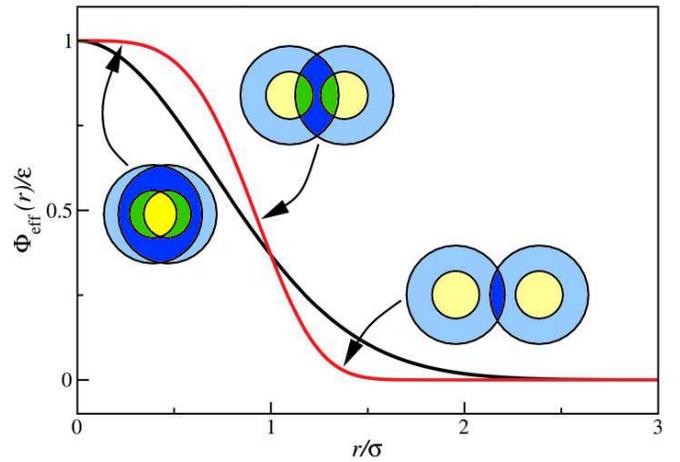}
\end{center}
\caption{(Color online) Schematic representation of the effective
potentials of amphiphilic dendrimers: interactions between the
different regions (solvophobic core - yellow, solvophilic shell -
blue) of amphiphilic dendrimers lead to interactions that are steeper
(red line) than the threshold of clustering (Gaussian, black line).}
\label{fig1}
\end{figure}

Following the ideas of material design, we modify the architecture of
athermal, flexible dendrimers along well-defined strategies. Since the
Gaussian effective interaction is at the threshold to clustering, we
require the following changes in $\Phi_{\rm eff}(r)$ to achieve
clustering behavior: a flatter core region, such as those of the
GEM-$n$ potentials with $n > 2$, which -- compared to a Gaussian --
also display a steeper decay of the repulsion for larger
separations. Alternatively, a positive effective interaction with a
local {\it minimum} at $r = 0$ also leads to oscillations in $\tilde
\Phi_{\rm eff}(q)$, as we have $\Phi_{\rm eff}''(r = 0) > 0$ in that
case.  To realize this goal, we aim for a more open structure and
stronger segregation between outer and inner particles by assembling
{\it amphiphilic} dendrimers built up from a solvophilic shell and
solvophobic core particles. In Fig.~\ref{fig1}, we qualitatively
illustrate how our modifications lead in the correct direction: as the
macromolecules start to overlap, the solvophilic and thus mutually
repulsive shells cause a steeply increasing potential wall.  This
effect is reinforced upon further decreasing the distance since core
and shell repel each other due to their different nature. Eventually,
the attractive core regions overlap and slow down further growth of
the repulsion, leading to a rather flat region or even a local minimum
in $\Phi_{\rm eff}(r)$ at small distances.

In an effort to realize these ideas we developed a computer model of
second generation \cite{ac:04} amphiphilic dendrimers
\cite{buzza:jcp:05} where the end-groups form the solvophilic shell
(index S) and all other monomers the solvophobic core (index C).  The
bonds between monomers are modeled by the finitely extensible
nonlinear elastic (FENE) potential \cite{Fene}
\begin{equation} \label{fene}
\beta \Phi_{\mu\nu}^{\rm FENE}(r) = - K_{\mu \nu} R_{\mu \nu}^2
\log\left[1-\left(\frac{r-l_{\mu \nu}^0}{R_{\mu \nu}}\right)^2\right],
\end{equation}
with $\mu\nu = {\rm CC,~CS}$,
which restricts the
bond-length to be in between $l_{\mu\nu}^{\rm max}$ and
$l_{\mu\nu}^{\rm min}$. $K_{\mu \nu}$ is the spring-constant and
$R_{\mu \nu}=l_{\mu\nu}^{\rm max}-l_{\mu \nu}^0$, with $l_{\mu
\nu}^0=(l_{\mu\nu}^{\rm max}+l_{\mu\nu}^{\rm min})/2$ being the
equilibrium bond length. All other interactions between two monomers
separated by distance $r$ are modeled by the Morse potential
\begin{equation} \label{morse}
\beta \Phi_{\mu \nu}^\mathrm{Morse}(r) = \varepsilon_{\mu \nu}
\left\{\left[e^{- \alpha_{\mu \nu}(r-d_{\mu
\nu})}-1\right]^2-1\right\},
\end{equation}
with $\mu\nu = {\rm CC,~CS,~SS}$, 
which is characterized by a repulsive core at short and an attractive
tail at long distances whose depth and range are parametrized by
$\varepsilon_{\mu \nu}$ and $\alpha_{\mu \nu}$, respectively. The
$d_{\mu \nu}$ are the monomer diameters. All potential parameters of
the dendrimers discussed in this work
are summarized in Table~\ref{table1}.

\begin{figure}
\begin{center}
\includegraphics[width=\columnwidth, clip=true, draft=false]
 {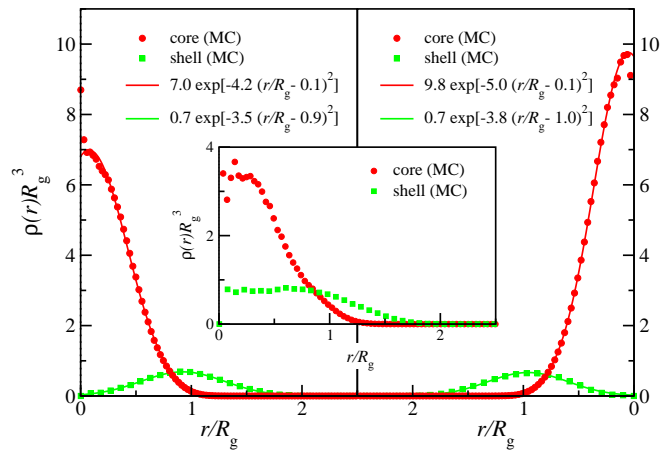}
\end{center}
\caption{(Color online) Monomer density profiles 
of the core (red) and the shell (green) region as
obtained by MC simulations (symbols) and from fits to Gaussians
(lines) for the amphiphilic dendrimers $D_1$ (left panel) and $D_2$
(right panel). The interaction parameters are given in Table I. In the
inset, we show the same for an athermal dendrimer of the same
generation.}
\label{gaussians}
\end{figure}

We calculated the monomer density profiles for the core and the shell
particles, $\rho_{\rm C}(r)$ and $\rho_{\rm S}(r)$, in standard Monte
Carlo (MC) simulations of an {\it isolated} dendrimer.  Representative
results pertaining to two ($D_1$, $D_2$) out of seven model dendrimers 
simulated are
shown in Fig.~\ref{gaussians}. They demonstrate that both the core- and
the shell-particle 
distributions are of Gaussian shape, with
the fitting parameters given 
in the respective panels. 
In striking contrast to athermal dendrimers 
(see Fig.\ \ref{gaussians}, inset),
these density profiles are -- due to amphiphilicity -- 
spatially segregated: the distribution of the core
monomers has its maximum close to the origin while the profile of the
shell particles is centered between $0.85\,R_g$ and $R_g$
(for all seven dendrimers simulated), with $R_g$ being
the dendrimer's radius of gyration. 

\begin{table}
  \begin{tabular}[t]{|c||c|c|c|}
    \hline
{\bf FENE} & $K/d_{\rm CC}^2$ & $l_0/d_{\rm CC}$ & $R/d_{\rm CC}$\\
 \hline \hline
CC  & 40 & 1.875  & 0.375 \\
\hline
CS  & 20 & 3.750  & 0.750 \\
\hline
  &  & 2.8125 ($D_1$) & 0.5625 ($D_1$) \\
\raisebox{2ex}[-2ex]{ZZ} &  \raisebox{2ex}[-2ex]{40} &  1.8750 ($D_2$) & 0.3750 ($D_2$) \\
    \hline  \hline
 {\bf  Morse} & $\varepsilon$ & $\alpha d_{\rm CC}$ & $d/d_{\rm CC}$\\
\hline  \hline
CC  & 0.714 &  6.4 &   1 \\ \hline
    &       &      & 1.50 ($D_1$) \\
\raisebox{2ex}[-2ex]{CS} &  \raisebox{2ex}[-2ex]{0.014} & \raisebox{2ex}[-2ex]{19.2}  & 1.25 ($D_2$) \\
\hline
  &  &  & 2.0 ($D_1$) \\
\raisebox{2ex}[-2ex]{SS} &  \raisebox{2ex}[-2ex]{0.014} & \raisebox{2ex}[-2ex]{19.2}  & 1.5 ($D_2$) \\
    \hline  
  \end{tabular}
\caption{Potential parameters of the dendrimers considered in this
study [cf.~Eqs.~(\ref{fene}) and (\ref{morse})], labeled $D_1$ and
$D_2$. ZZ refers to the two central monomers.}
\label{table1}
\end{table}

Next, we performed MC simulations between two interacting dendrimers,
averaging over the degrees of freedom of the constituent monomers to
determine the effective potential, $\Phi_{\rm eff}(r)$. Allowing two
dendrimers to interact freely, their effective potential can be
determined from the centers-of-mass (c.o.m.)~correlation function $G(r)$, given by
$G(|{\bf R}_1-{\bf R}_2|) = \langle \hat \varrho_1({\bf R}_1)\hat
\varrho_2({\bf R}_2)\rangle$. Here, 
$\hat \varrho_i({\bf R}_i) = \delta({\bf R}_i - {\bf S}_i)$
is the density operator for the c.o.m. of dendrimer $i$,
and the average $\langle\cdots\rangle$
is taken over all c.o.m.\ position vectors ${\bf S}_i$ ($i=1,2$).
$\Phi_{\rm eff}(r)$ is then given by $\beta \Phi_{\rm eff}(r) = -
\ln[G(r)]$, with $\beta = 1/k_{\rm B}T$. Since the repulsion between
the dendrimers is expected to be strong at short distances, this
scheme will provide only poor statistics for small separations. To
cope with this problem, we use non-Boltzmann sampling
\cite{chandler:87} where we divide the $r$-range into $m=15$
windows of width $\Delta r_j = \frac{2 r_{\rm max}}{m (m+1)}(j+1)$ and assume for each window $j=0,\dots,m-1$ an
umbrella potential $W_j(r)$:
\begin{equation}
W_j(r) = \left\{ \begin{array}{ll} 
                0 & r_j -\delta r < r < r_j + \Delta r_j + \delta r \\ 
                \infty & \mathrm{else} \end{array} \right. .
\end{equation}
Here, $r_0=0$ and else $r_j = \sum_{i=0}^{j-1} \Delta r_i$. Further, $r_{\rm max}= 5
R_g$ and $\delta r$ is chosen to guarantee for slightly
overlapping windows (at the edges, $\delta r=0$.) 
For each of these windows,
simulations of $2 \times 10^8$ MC sweeps are performed and $G_j(r)$ is
determined to within an additive constant due to normalization. We
obtain $\Phi_{{\rm eff},j}(r)$ in each window $j$ as $-k_{\rm
B}T\ln\,G_j(r)$ and these results are merged to form a continuous
curve which is finally normalized by setting $\beta \Phi_{\rm
eff}(r=r_\mathrm{max})=0$.

\begin{figure}
\begin{center}
\includegraphics[width=\columnwidth, clip=true, draft=false]
 {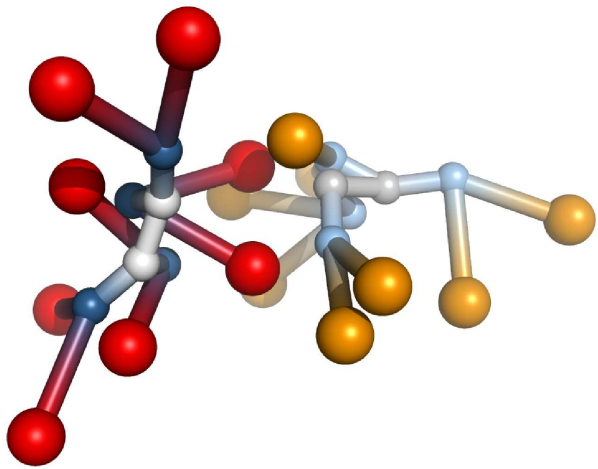}
\end{center}
\caption{(Color online) Simulation snapshot of two interacting amphiphilic
dendrimers, both showing a dense shell conformation.}
\label{fig3}
\end{figure}

\begin{figure}
\begin{center}
\includegraphics[width=\columnwidth, clip=true, draft=false]
 {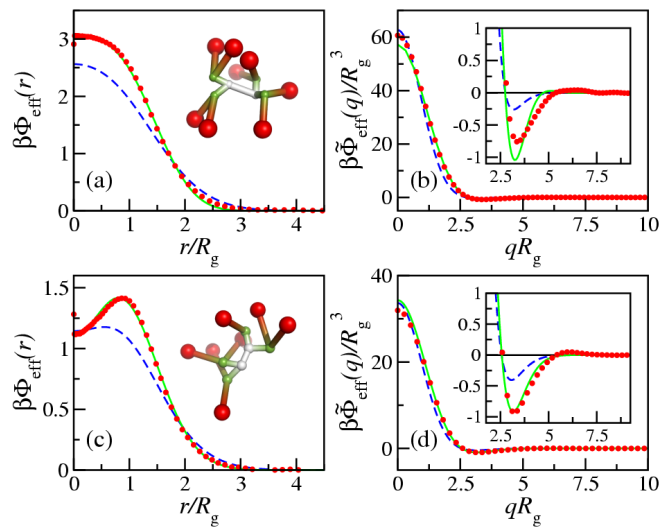}
\end{center}
\caption{(Color online) The effective potentials $\Phi_{\rm eff}(r)$ of
typical amphiphilic dendrimers [(a), (c)] and their FTs
$\tilde \Phi_{\rm eff}(q)$ [(b), (d)], showing negative parts.
The blue dashed line denotes the theoretical result (see text),
and the green lines are fits to the simulation data.
(a) and (b) pertain to dendrimer $D_1$, (c) and (d) to $D_2$.
The insets in (a)
and (c) feature simulation snapshots of the respective dendrimers.}
\label{fig4}
\end{figure}
 
A typical simulation snapshot of two interacting dendrimers is shown
in Fig.~\ref{fig3}. Results for the effective interactions of the two
dendrimers are summarized in Fig.~\ref{fig4}.  $\Phi_{\rm eff}(r)$
indeed shows a steep increase as the macromolecules approach and
eventually becomes rather flat [Fig.~4(a)], or even exhibits a
locally attractive dip [Fig.~4(c)] for smaller distances. 
We fitted $\Phi_{\rm eff}(r)$ of dendrimer $D_1$ by a GEM-$n$
potential, finding that $\Phi_n(r)$ with an index $n \cong 3.1$ 
approximates it
with high accuracy.  The effective potential of
dendrimer $D_2$ can be fitted to a double-Gauss interaction of the
form $\Phi_{\rm eff}(r) = \epsilon_1\exp[-(r/\sigma_1)^2] -
\epsilon_2\exp[-(r/\sigma_2)^2]$, where $\epsilon_1 = 23.6 k_{\rm B}T
$, $\epsilon_2 = 22.5 k_{\rm B}T$, $\sigma_1 = 1.117 R_g$ and
$\sigma_2 = 1.059 R_g$.  Both effective potentials lead to
clustering, since $\Phi_{\rm eff}''(r=0) \geq 0$. 

To provide a semi-quantitative theoretical background to these
simulation results, we reconsider our amphiphilic dendrimers within a
suitably modified Flory theory. The original idea of this concept
\cite{doi:03} is based on the simplifying assumptions that the 
spherosymmetric monomer
densities of isolated athermal dendrimers
around the c.o.m., $\rho(r)$, do not change
upon close interaction. 
Here, $\rho(r) = \langle\sum_j\delta({\bf r}-{\bf r}_j)\rangle$
and ${\bf r}_j$ denotes the position vector of monomer $j$ with 
respect to the center of mass.
Thus, considering two such dendrimers at separation
${\bf R}$ and {\it assuming} that their profiles are not
distorted by their mutual presence, the effective interaction between
the two dendrimers takes the form
\begin{equation}
\Phi_{\rm eff}(|{\bf R}|) = \iint \rho(|{\bf r}_1|)\rho(|{\bf r}_2-{\bf R}|)
v(|{\bf r}_1-{\bf r}_2|){\rm d}{\bf r}_1{\rm d}{\bf r}_2,
\label{flory1:eq}
\end{equation}
where $v(|{\bf r}_1-{\bf r}_2|)$ is the monomer-monomer
interaction. For the latter, a contact interaction weighted by
the second virial coefficient $v_0$ of the monomer-monomer interaction
is introduced, 
$\beta v(|{\bf
r}_1-{\bf r}_2|) = v_0\delta(|{\bf r}_1-{\bf r}_2|)$. For athermal
dendrimers, $v_0 > 0$. The FT of $\Phi_{\rm eff}(r)$ given in
Eq.~(\ref{flory1:eq}) then reads in this case $\beta \tilde \Phi_{\rm eff}(k) =
v_0\tilde\rho^2(k)>0\,\,\forall k$.

Generalizing this model to amphiphilic dendrimers, we treat the
core and the shell profiles separately, introducing {\it three}
different 
excluded volume parameters, $v_{\rm CC}$, $v_{\rm CS}$, and $v_{\rm
SS}$ given by the second virial coefficients of the underlying 
monomer-monomer interactions.
Proceeding along similar lines as above leads to the FT of the
effective interaction between amphiphilic dendrimers as
$\beta \tilde \Phi_{\rm eff}(k) = \sum_{\mu,\nu} v_{\mu\nu}
 \tilde\rho_{\mu}(k)\tilde\rho_{\nu}(k)$ 
where $\tilde\rho_{\mu}(k)$ is the FT of 
$\rho_{\mu}(r)$ and $\mu, \nu = {\rm C,~S}$. Core solvophobicity implies
$v_{\rm{CC}}<0$, whereas $v_{\rm CS}, v_{\rm SS} > 0$.  
Consequently, $\tilde \Phi_{\rm eff}(k)$ can also display
negative components. 
The values of the 
second virial coefficients for our dendrimers are
$v_{\rm CC} = -1.90$, $v_{\rm SS} = 28.40$ and 
$v_{\rm CS} = 11.31$ ($D_1$), and
$v_{\rm CC} = -1.90$, $v_{\rm SS} = 11.31$ and
$v_{\rm CS} = 6.25$ ($D_2$), measured in 
$d_{\rm CC}^{-3}$.
Based on the simulation results
(cf.~Fig.~\ref{gaussians}), we model the monomer densities
$\rho_\mu(r)$ as Gaussian functions,
$\rho_\mu(r) = S_\mu\exp[-\gamma_\mu (r-r_\mu)^2]$,
taking for $S_\mu$, $\gamma_\mu$, and $r_\mu$, $\mu= {\rm C,~S}$, those
values that provide the best fit of the simulation data, quoted in 
Fig.~\ref{gaussians}. Approximate expressions for the $\tilde
\rho_{\mu}(k)$ are given in \cite{HanXX}. 
The theoretical results for $\tilde \Phi_{\rm eff}(k)$ and hence
$\Phi_{\rm eff}(r)$ are shown
in Fig.~\ref{fig4} along with the data extracted from the
simulations. In view of the simplifying assumptions of Flory theory,
the good qualitative agreement between simulations and theory is 
astonishing. The negative Fourier components
[Figs.~\ref{fig4}(b) and \ref{fig4}(d)],
are less pronounced in theory than in simulation,
thus the former provides a lower threshold to the
onset of clustering.


The counterintuitive phenomenon of clustering in the complete absence
of attraction might motivate experimental groups to assemble
amphiphilic dendrimers in the lab. To this end let us summarize our
guidelines for synthesizing clustering dendrimers.  In a first step,
suitable solvophobic core and solvophilic shell groups have to be
chosen for the experiments, for which simulations on an isolated
dendrimer are performed, leading to the core- and the shell-density
profiles. While Flory theory provides a reliable qualitative indicator
whether the threshold to clustering has already been reached, full
evidence can then be gathered by measuring the effective interactions
in the more time-consuming simulations of two interacting
dendrimers. From our observations for all dendrimers investigated, the
following remarks should prove valuable: bigger end-groups and/or
shorter end-group spacers lead to a stronger repulsion in $\Phi_{\rm
eff}(r)$ at short distances. Further, if the spacer length of the
end-groups is increased and/or the distance 
between the two central particles and/or the end-group particle size is
reduced, $\Phi_{\rm eff}(r)$ becomes flatter or even develops a dip at
small distances.  
The low dendrimer-generation number 
is encouraging for experimentalists
because it hints at a rather straightforward synthesis process
\cite{antoni:07}.  Since $\Phi_{\rm eff}(r=0) \sim k_{\rm B} T$,
clustering can easily be realized under ambient conditions in
thermally activated processes.

The findings of this work bear significance for soft matter science
and materials design at various levels. At the one-particle level, we
have established that synthesizing open dendrimers with a segregated
core-shell structure requires neither stiff bonds nor electrostatic
repulsions as commonly believed: amphiphilicity is sufficient.  At the
many-body level, solutions of such dendrimers will display pronounced
correlations at a {\it single} length scale, independently of the
density \cite{cnl:jcp:07}, allowing thus for well-controlled spatial
modulation of confined liquids and, e.g., their local index of
refraction, whose intensity can be tuned by changing the degree of
confinement. Crystals formed by such systems show density-{\it
independent} lattice constants, a novel of
self-assembly of condensed matter. Such crystals are
quite unusual, since they are diffusive on the single-particle level,
allowing thus mass transport but arrested, and thus rigid as a
conventional solid, at the collective level \cite{angel:prl:07}.
Finally, on the fundamental level, we have demonstrated that within
soft matter, bounded effective interactions can be manipulated with
the same degree of flexibility as diverging ones do. 

We thank D.~Gottwald for helpful discussions. This work
was funded by the FWF, Proj.~No.~17823-N08 and by the
DFG within the SFB-TR6, Project Section C3.

\end{document}